# Water as a Battery: Using Hydropower Flexibility to Support the Grid

**C. MORANKAR, A. AL MEHADI***
**WSP USA, Inc.**
**USA**

## SUMMARY

This study presents a two-layer optimization framework to determine the optimal sizing of PV, wind, and battery storage co-located with existing hydroelectric plants to support rapidly increasing data center loads while minimizing cost, grid reliance, and transmission upgrades. The outer layer uses Particle Swarm Optimization (PSO) to select capacity combinations, while the inner layer performs rolling-horizon dispatch simulations to model hourly energy balance among PV, wind, hydro, and BESS, considering operational constraints. Two case studies were analyzed. At The Dalles Dam (Oregon), favorable wind resources and land availability enabled deployment of 111.2 MW PV and 117.9 MW wind, with no BESS required. Hydro flexibility allowed the system to meet the load with minimal grid support. In contrast, Conowingo Dam (Maryland) faced land constraints and limited renewables to 130 MW PV, leading to higher grid dependence. Co-locating hydro with renewables significantly enhances reliability and cost effectiveness by leveraging hydro's dispatchability to balance daily and seasonal variable generation from PV+Wind, reducing the need for storage or costly grid imports. This analysis provides a framework to meet the 99.9% availability targets of this new load type on the grid with sustainable energy resources.

## INTRODUCTION

Achieving global climate goals and net-zero emissions by mid-century will require an unprecedented expansion of renewable energy, accompanied by a corresponding increase in grid flexibility. The transition to wind and solar power is already underway at a record pace. The International Energy Agency projects that these variable resources could supply 54–71% of global electricity by 2050, a scenario critical for reaching net-zero targets [1]. However, such high renewable penetration poses challenges due to the intermittent nature of solar and wind; without sufficient energy storage or flexible generation to buffer this variability, power grids

abdullah.mehadi@usu.edu

risk imbalances where supply overshoots or falls short of demand. Ensuring reliability in a renewable-dominated grid thus demands new strategies to provide fast-balancing capacity and store excess energy when needed. In parallel, the energy demand of data centers – the digital infrastructure underpinning cloud services and AI is soaring and increasingly under scrutiny [3]. In the United States alone, data centers consumed roughly 176 TWh in 2023 (≈4.4% of national electricity use), and their annual consumption could grow to 1,050 TWh (∼12% of U.S. demand) by 2030 [2]. Globally, electricity use by data centers is on track to more than double to ~945 TWh by 2030, equivalent to the entire power consumption of Japan. This rapid growth, if met with fossil-based power, would hinder climate objectives – currently about 56% of U.S. data center electricity still comes from fossil fuels. In response, major technology firms and hyperscale data center operators are exploring reliable options to power their facilities with 100% carbon-free energy.

These trends underscore a pressing need for solutions that can reliably meet large, round-the clock loads like data centers using mostly renewable energy. One promising approach is to leverage hydropower's flexibility as a "water battery" to integrate renewable generation and supply such constant loads. Conventional and pumped-storage hydropower plants can rapidly adjust output and effectively store energy by controlling water flow, making them an ideal partner to balance variable solar and wind output. By co-locating new solar PV and wind farms with existing hydroelectric dams, the excess renewable energy can be "stored" in the form of water in the reservoir (analogous to charging a battery) and released through turbines when solar/wind production wanes or when demand spikes. This synergy can minimize reliance on backup fossil generators or large battery banks, instead using the dam's dispatchable capacity to smooth out fluctuations. Such hybrid renewable-hydro systems align well with the needs of data centers by offering a sustainable, highly reliable power supply: the intermittent solar and wind generation is buffered by the hydro plant's on-demand output, ensuring continuous delivery of electricity even when "the wind doesn't blow and the sun doesn't shine." Moreover, it also eliminates or reduces the need for traditional BESS. This hybrid approach also allows us to increase the capacity factor of the existing substation infrastructure at the hydro power facility. Combining PV+Wind with existing hydro substation can help increase the capacity factor from 40-50% to 80%-90%.

In this study, we develop a two-layer optimization framework to design and assess a co-located renewable energy system with hydropower for data center applications. The framework determines the optimal sizes of on-site solar PV, wind, and battery energy storage systems (BESS) at an existing hydroelectric site, aiming to meet a given data center load at minimum cost and with minimal grid draw. An outer-layer optimization using Particle Swarm Optimization (PSO) selects candidate capacity combinations, while an inner-layer simulation performs rolling-horizon dispatch of PV, wind, hydro, and BESS on an hourly basis. This nested approach captures the operational constraints of the hydro plant (e.g. water discharge limits) and the variability of renewable generation, ensuring the proposed mix can feasibly balance the data center's demand every hour.

We apply this methodology to two contrasting case studies: The Dalles Dam in Oregon and Conowingo Dam in Maryland. The Dalles case has favorable wind resources and available land. In contrast, the Conowingo case faces land and resource limitations that restrict on-site renewables. The next two sections discuss the methodology and the results.

METHODOLOGY

| Variable | Description |
|---|---|
| $C_{\text{PV}}$ (MW) | Installed capacity of PV system |
| $C_{\text{Wind}}$ (MW) | Installed capacity of wind system |
| $E_{\text{BESS}}$ (MWh) | Energy capacity of the battery storage system |
| $G_{\text{PV}}(t)$ (MW) | PV generation at time $t$ |
| $G_{\text{Wind}}(t)$ (MW) | Wind generation at time $t$ |
| $H_{\text{saved}}(t)$ (MWh) | Amount of hydro energy preserved (not dispatched) at time $t$ for future use. |
| $s_{\text{PV}}(t)$ (p.u.) | Normalized PV generation profile from SAM |
| $s_{\text{Wind}}(t)$ (p.u.) | Normalized wind generation profile from SAM |
| $\eta_{\text{inv}}$ | PV inverter/DC–AC conversion efficiency |
| $\eta_{\text{turb}}$ | Wind turbine electrical efficiency |
| $\eta_{\text{conv}}$ | Generic converter efficiency (if applied) |
| $L_t$ (MW) | Load demand at time $t$ |
| SOC (p.u.) | Battery state-of-charge |
| $\mathbf{x}_i$ | Position of particle $i$ in PSO |
| $\mathbf{v}_i$ | Velocity of particle $i$ in PSO |
| $\mathbf{p}_i$ | Personal best position of particle $i$ |
| $\mathbf{g}$ | Global best position across all particles |
| $c_1$, $c_2$ | Acceleration constants in PSO |
| $r_1$, $r_2$ | Random numbers in $[0,1]$ |
| $w$ | Inertia weight in PSO |
| $E_{\text{PV}}$ (MWh) | Total PV energy output over simulation horizon |
| $E_{\text{BESS}}$ (MWh) | Total BESS energy discharged over horizon |
| $E_{\text{Wind}}$ (MWh) | Total wind energy output over horizon |
| $Cost_{\text{BESS}}$ ($/MWh) | Unit cost of BESS including CAPEX, O&M, replacements |
| $Cost_{\text{PV}}$ ($/MWh) | Unit cost of PV including CAPEX, O&M |
| $Cost_{\text{Wind}}$ ($/MWh) | Unit cost of wind system |
| $Cost_{\text{Grid}}$ ($) | Cost of imported grid electricity |
| $Cost_{\text{Degradation}}$ ($) | Cost of BESS degradation due to cycling |
| $Cost_{\text{Ramp}}$ ($) | Penalty cost for ramp rate violations |
| $EC$ ($) | Total energy cost from PV, wind, and BESS |
| $Cost_{\text{PV,daily}}$ | Daily PV operational cost |
| $Cost_{\text{Wind,daily}}$ | Daily Wind operational cost |
| $Cost_{\text{BESS,daily}}$ | Daily BESS operational cost |
| $T_{\text{operation}}$ | Number of operational days of the plant |
| $T_{\text{BESS}}$ | Number of operational days of the BESS |

Table 1: Summary of variables and parameters used in the optimization framework.

This study proposes a two-layer optimization framework to determine optimal capacities of photovoltaic (PV), wind, and battery energy storage system (BESS) to support a low-carbon power system integrated with an existing hydroelectric plant. The algorithm combines a Particle Swarm Optimization (PSO) technique with a rolling-horizon dispatch model for short-term planning.

## Outer Layer: Particle Swarm Optimization

The Particle Swarm Optimization (PSO) layer searches the decision space for the optimal combination of: PV capacity ($C_{PV}$ in MW), Wind capacity ($C_{Wind}$ in MW), BESS energy capacity ($E_{BESS}$ in MWh).

Each particle $\mathbf{x}i = [C_{PV}, C_{Wind}, E_{BESS}]$ evolves over iterations by updating its velocity and position according to:

$$\mathbf{v}_i^{(t+1)} = w \cdot \mathbf{v}_i^{(t)} + c_1 r_1 (\mathbf{p}_i - \mathbf{x}_i^{(t)}) + c_2 r_2 (\mathbf{g} - \mathbf{x}_i^{(t)})\ \mathbf{x}_i^{(t+1)} \qquad = \mathbf{x}_i^{(t)} + \mathbf{v}_i^{(t+1)}$$

Where $\mathbf{p}_i$ is the personal best of particle $i$, $\mathbf{g}$ is the global best, $r_1$ and $r_2$ are random numbers in $[0,1]$, and $c_1$, $c_2$ are acceleration constants.

The objective is to minimize the total cost as evaluated by the dispatch simulation described below.

## Inner Layer: Rolling-Horizon Dispatch

The dispatch simulator operates on an hourly resolution and computes the contribution of PV, wind, hydro, and BESS to meet the load profile. At each time step $t$: 1. Compute hourly PV and wind generation:

$$G_{PV}(t) = C_{PV} \cdot \eta_{conv} \cdot s_{PV}(t)$$

$$G_{Wind}(t) = C_{Wind} \cdot \eta_{conv} \cdot s_{Wind}(t)$$

$$H_{saved}(t) \geq 0$$

The dispatch strategy is described as follows: If $G_{PV}(t) + G_{Wind}(t) \geq L_t$, use renewables to meet the load, defer hydro, and charge the battery with surplus energy. Else, dispatch hydro (including stored water), and discharge the battery if needed. Grid power is only used when the maximum transformer capacity is exceeded. Update battery state-of-charge (SOC), hydro storage, and enforce ramp rate and SOC constraints. Fig. 1 shows the simple flowchart of the hydro dispatch strategy where $C_{saved}(t)$ is the saved hydro power which can be any number greater than 0.

## Cost Model

The total energy cost (EC) is the objective function and is computed as :

$$EC = E_{PV} \cdot Cost_{PV} + E_{BESS} \cdot Cost_{BESS} + E_{Wind} \cdot Cost_{Wind}$$

Here, $E_{PV}$ and $E_{BESS}$ are the total output energy of PV and BESS over the simulation horizon. $C_{PV}$, $C_{Wind}$ represents unit cost (\$/MW) for PV and Wind while $C_{BESS}$ represent the unit costs (\$/MWh) of BESS respectively, including capital, replacement, and O&M costs. Additional charges include:

$$\text{Total Cost} = EC + Cost_{\text{Grid}} + Cost_{\text{Degradation}} + Cost_{\text{Ramp}}$$

where, $Cost_{\text{Grid}}$ is the energy imported from the grid at time-varying rates, $Cost_{\text{Degradation}}$ is proportional to battery throughput and $Cost_{\text{Ramp}}$: penalizes rapid changes in BESS output.

Infeasible solutions where the demand is unmet result in a large penalty (e.g., $10^7$), thus avoiding infeasible solution that doesn't maintain the constraints.

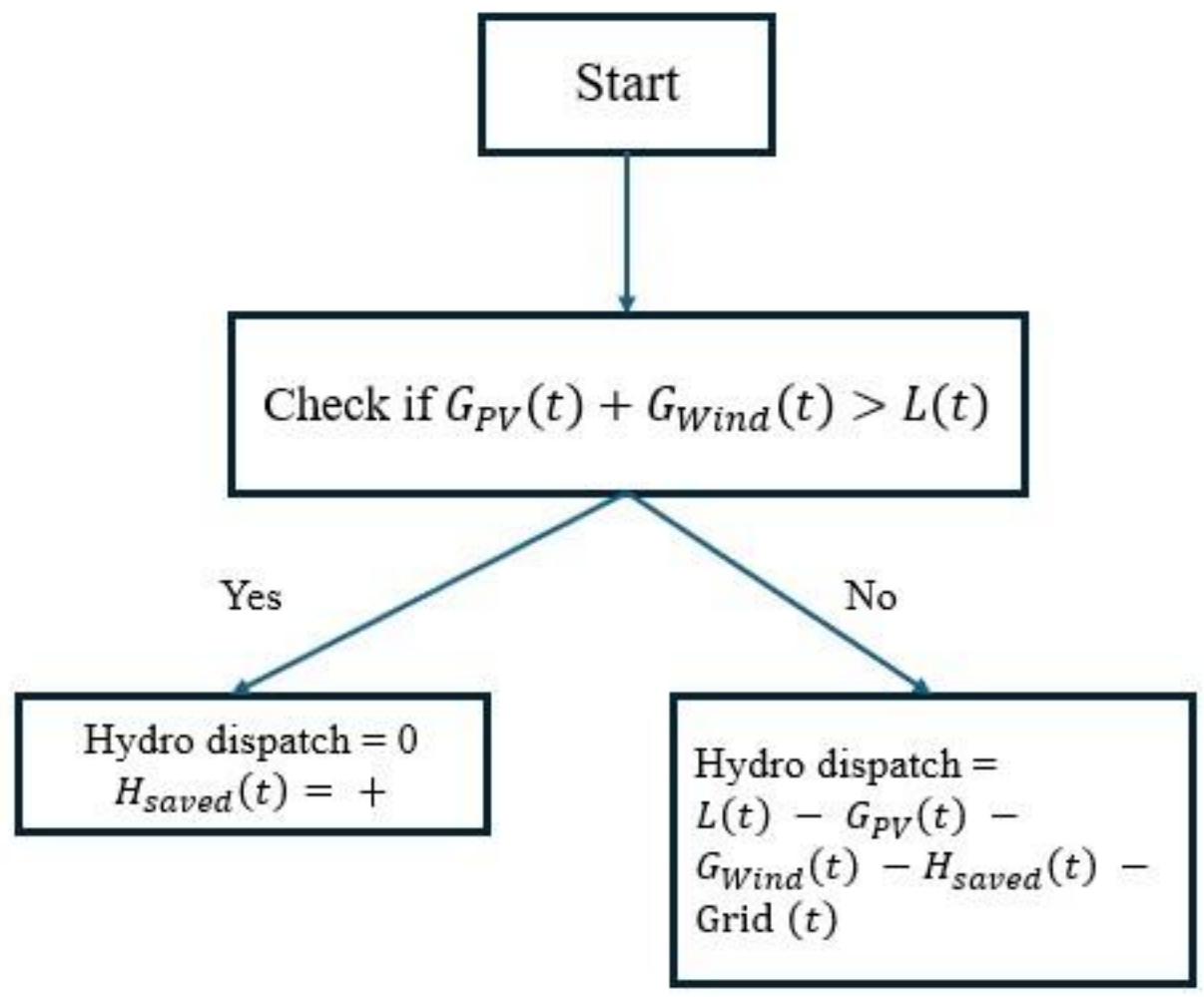

Figure 1: Hydro dispatch flowchart

## PV and Wind Generation

For the base PV generation and wind generation profile, NREL's System Advisor Model (SAM) was used, leveraging Typical Meteorological Year (TMY) weather data from the corresponding location. SAM provides detailed sub-hourly simulations of renewable energy output based on physical parameters, system configurations, and site-specific weather inputs.

The hourly generation output for PV and wind was extracted from SAM for a representative day, and scaled in the optimization as:

$$G_{\text{PV}}(t) = C_{\text{PV}} \cdot \eta_{\text{inv}} \cdot s_{\text{PV}}(t)$$

$$G_{\text{Wind}}(t) = C_{\text{Wind}} \cdot \eta_{\text{turb}} \cdot s_{\text{Wind}}(t)$$

Here: $G_{\text{PV}}(t)$ and $G_{\text{Wind}}(t)$ are the net AC power outputs (in MW) at time $t$. $C_{\text{PV}}$ and $C_{\text{Wind}}$ are the installed capacities (in MW) of PV and wind. $\eta_{\text{inv}}$ is the inverter and DC-AC conversion efficiency for PV. $\eta_{\text{turb}}$ is the electrical efficiency of the wind turbine system. $s_{\text{PV}}(t)$ and $s_{\text{Wind}}(t)$ are the normalized generation profiles (per unit output from SAM). Moreover, The normalized SAM output $s(t)$ was used as a template profile, and during simulation, actual capacity values were applied as decision variables. The generation data reflects hourly irradiance, temperature, wind speed, and other atmospheric parameters impacting system output.

## SIMULATION ENVIRONMENT

### PV System

At the geographic location of The Dalles Dam (45.6111 N, 121.1241 W) and Conowingo Dam (39.6563 N, 76.1758 W), a representative photovoltaic (PV) system was modeled using NREL's System Advisor Model (SAM) to generate hourly solar generation profiles. The simulation assumed a 1 MW nameplate capacity system was used with DC-to-AC ratio of 1.2. This configuration aligns with typical utility-scale deployment standards and captures both direct and albedo irradiance effects, which are particularly important for bifacial technologies. The NSRDB (National Solar Radiation Database) Typical Meteorological Year (TMY) weather file for the site was used to ensure realistic irradiance and temperature conditions. The resulting hourly AC power output from SAM served as the base PV profile input for the optimization framework.

### Wind System

For the wind energy system single wind turbine at hub height of 100 meter was simulated using the wind energy resource data available in SAM for the location Dalles Dam (45.6111 N, 121.1241 W). The parameters were kept simple and simulated. Thus, the resulting hourly AC wind power output from SAM served as the base Wind profile input for the optimization framework which was normalized by dividing by highest generation amount.

### Component Cost Assumptions

The total cost of the hybrid energy system includes capital costs, operational costs, and performance-related penalties. This section details the calculation of the component-wise capital costs and throughput degradation cost for the battery. The capital cost for PV is assumed to be a fixed daily cost per unit capacity installed:

$$Cost_{\text{PV}} = Cost_{\text{PV,daily}} \cdot C_{\text{PV,MW}} \cdot T_{\text{operation}}$$

where: $Cost_{\text{PV,daily}}$ is the per day CAPEX for PV $/MW/day, $C_{\text{PV,MW}}$ is the installed PV capacity (MW), $T_{\text{operation}}$ is the number of operational days of the plant. The wind capital cost is similarly computed:

$$Cost_{\text{Wind}} = Cost_{\text{Wind,daily}} \cdot C_{\text{Wind,MW}} \cdot T_{\text{operation}}$$

where $C_{\text{Wind,daily}}$ denotes the aggregated per day cost for wind per MW. On the other hand, the BESS capital cost is modeled as:

$$Cost_{\text{BESS}} = Cost_{\text{BESS,daily}} \cdot E_{\text{BESS}} \cdot T_{\text{bess}}$$

where: $C_{\text{BESS,daily}}$ is the per day cost per MWh of BESS, $E_{\text{BESS}}$ is the installed energy capacity of BESS (MWh). $T_{\text{operation}}$ is the operational period of the PV and Wind plant which is assumed to be 20 years : (20 × 365 = 7300 days) while $T_{\text{bess}}$ is the economic life in days : (15 years × 365 = 5475 days).

To account for degradation, a throughput-based cost is added based on energy cycled through the battery:

$$Cost_{\text{Deg}} = R_{\text{deg}} \cdot \sum_t |P_{\text{BESS}}(t)| \cdot \Delta t$$

where: $R_{\text{deg}}$ is the degradation cost rate (\$/MWh), $P_{\text{BESS}}(t)$ is the power charged or discharged at time $t$, $\Delta t$ is the time step in hours (typically 1).

Hence, the total investment cost is the sum of all above components:

$$Cost_{\text{Total}} = Cost_{\text{PV}} + Cost_{\text{Wind}} + Cost_{\text{BESS}} + Cost_{\text{Deg}} + Cost_{\text{O\&M}} + Cost_{\text{Grid}} + Cost_{\text{Penalty}}$$

This cost is used as the objective function in the outer-layer PSO to determine optimal PV, wind, and BESS sizing.

## Results

### Data center load

In our case it was considered that 55 MW of IT load of the data center will be added to each site. Thus accounting for 1.4 PUE (power usage efficiency) the peak demand is 77 MW and 24 hour energy requirement 1848 MWh.

### Case-1 The Dalles hydro power plant at Oregon

In Oregon, where wind already dominates the generation mix due to proximity to the coast and the availability of large land areas near the Dalles Dam, the optimization yielded an optimal PV capacity of 111.2 MW and wind capacity of 117.9 MW. Figure 2 and Figure 3 illustrate the PV and wind profiles for the Oregon region. The generation profile at the Dalles Dam is shown in Figure 4, while Figure 5 compares the hydro output before and after optimization. Figure 6 presents the stacked generation plot of the entire system, incorporating both PV and wind resources. An infrastructure/transformer capacity limit of 854.21 MW was considered. Due to the operational flexibility of the hydro plant, no battery energy storage system (BESS) was required. The cost breakdown of the system is provided in Table 2. Thanks to the significant contribution from PV and wind generation, grid reliance was minimal, as indicated by the red area in Figure 6, primarily driven by the high renewable output.

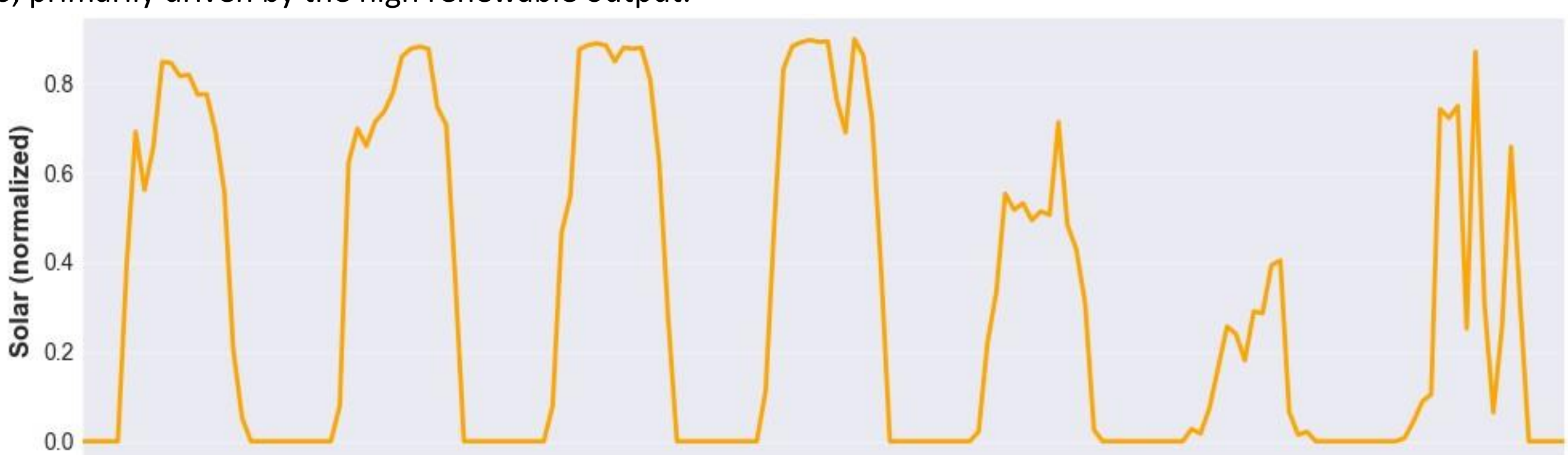

Figure 2: 1 MW solar PV profile at Dalles Dam normalized

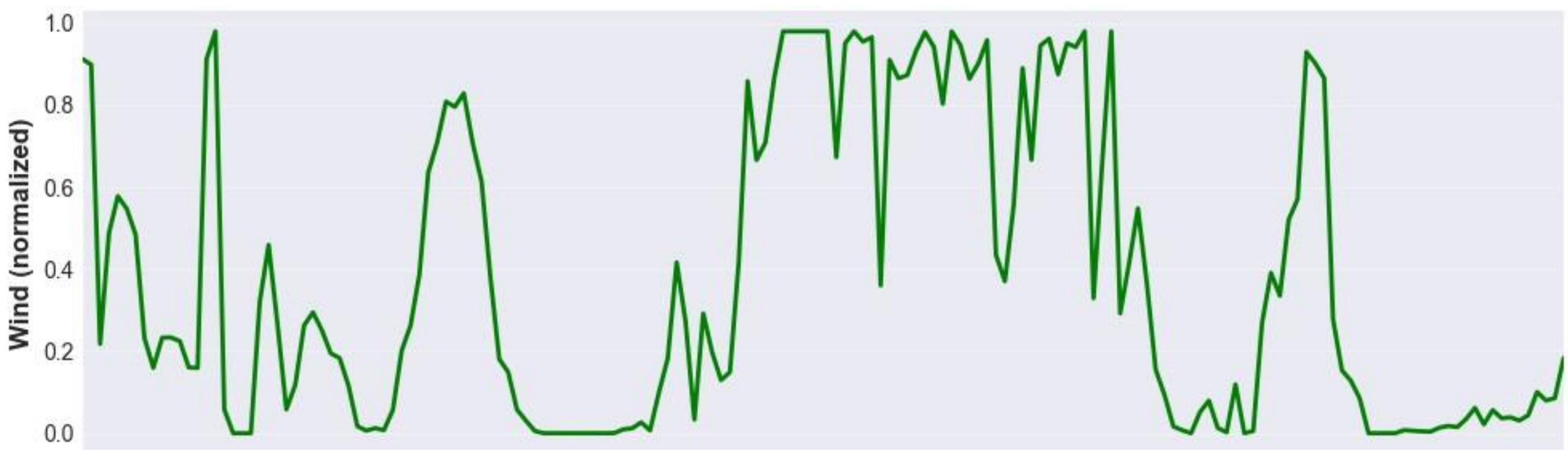

Figure 3: 1 MW wind profile at Dalles Dam normalized

Table 2: System Cost Breakdown at Dalles Dam

| Component | Cost (USD) |
|---|---|
| PV CapEx + O&M | $187,509,478.86 |
| Wind CapEx + O&M | $243,491,920.31 |
| BESS CapEx + O&M | $0.00 |
| Grid Cost | $288,302.07 |
| Total Cost | $187,797,780.93 |

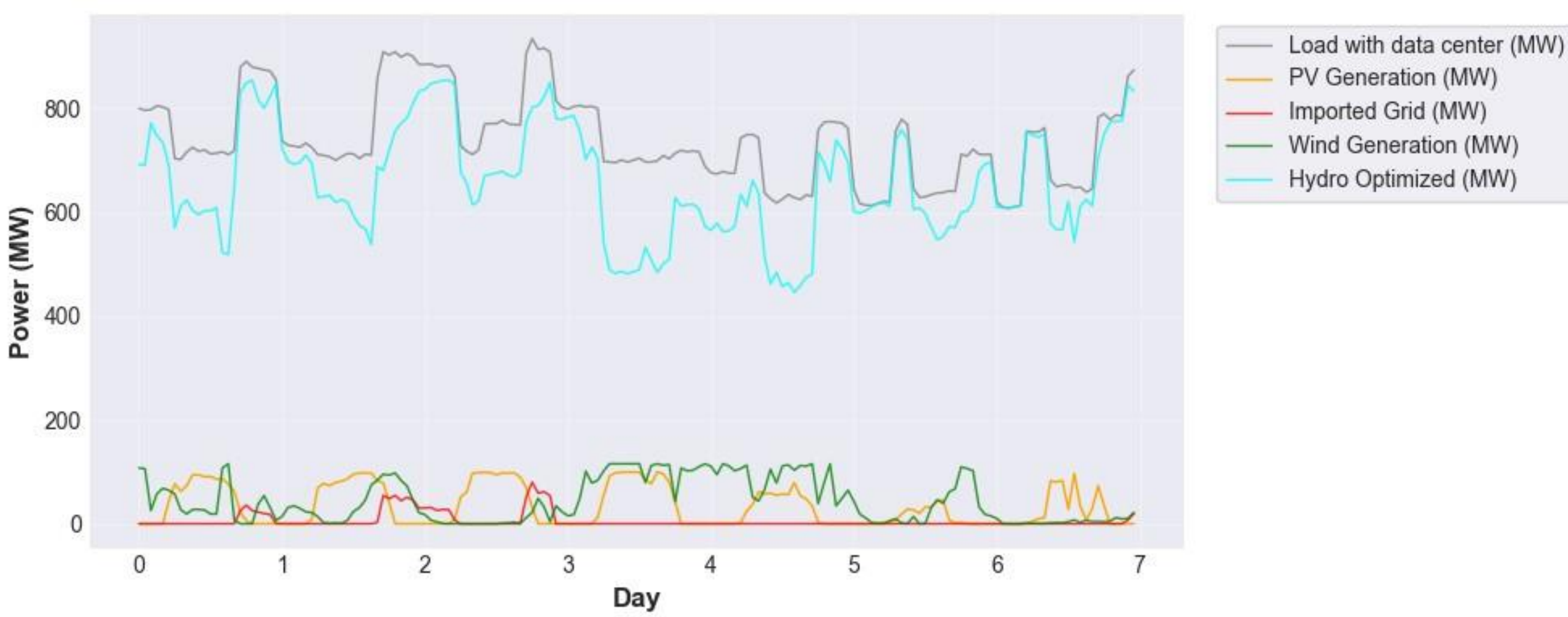

Figure 4: 7 days profile at Dalles Dam

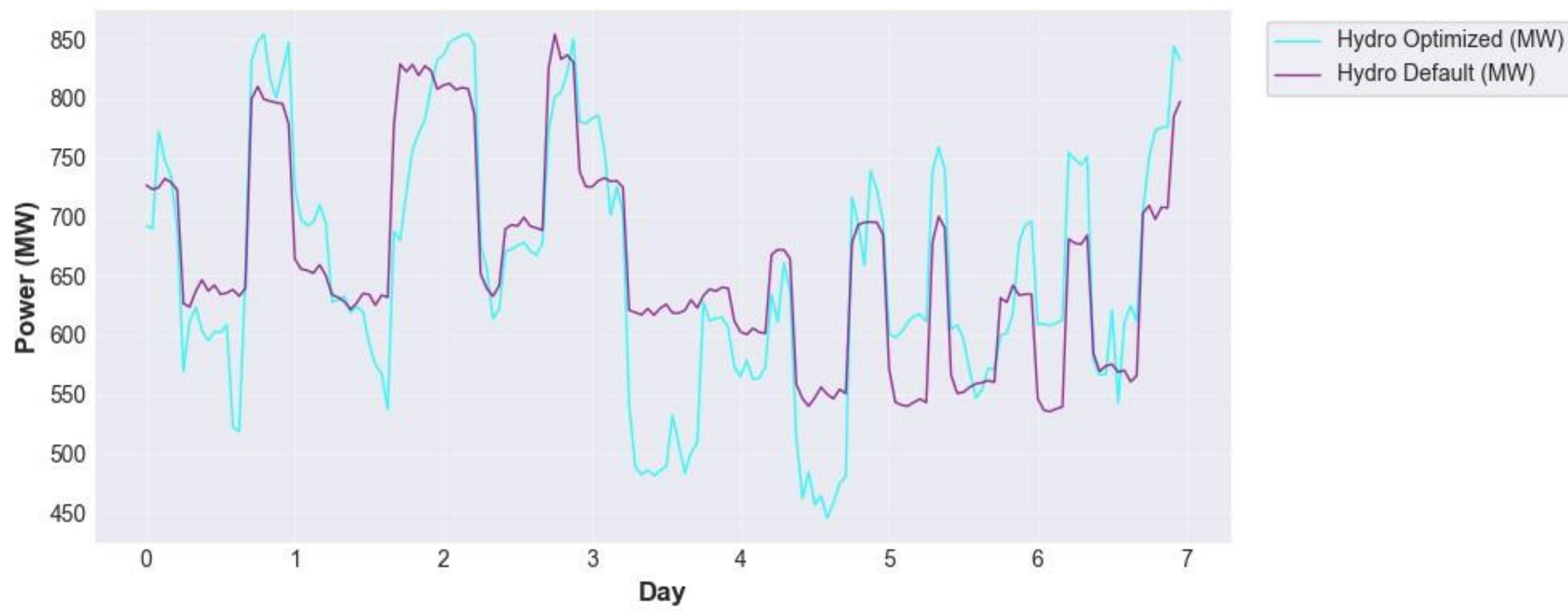

Figure 5: 7 days optimized hydro profile at Dalles Dam

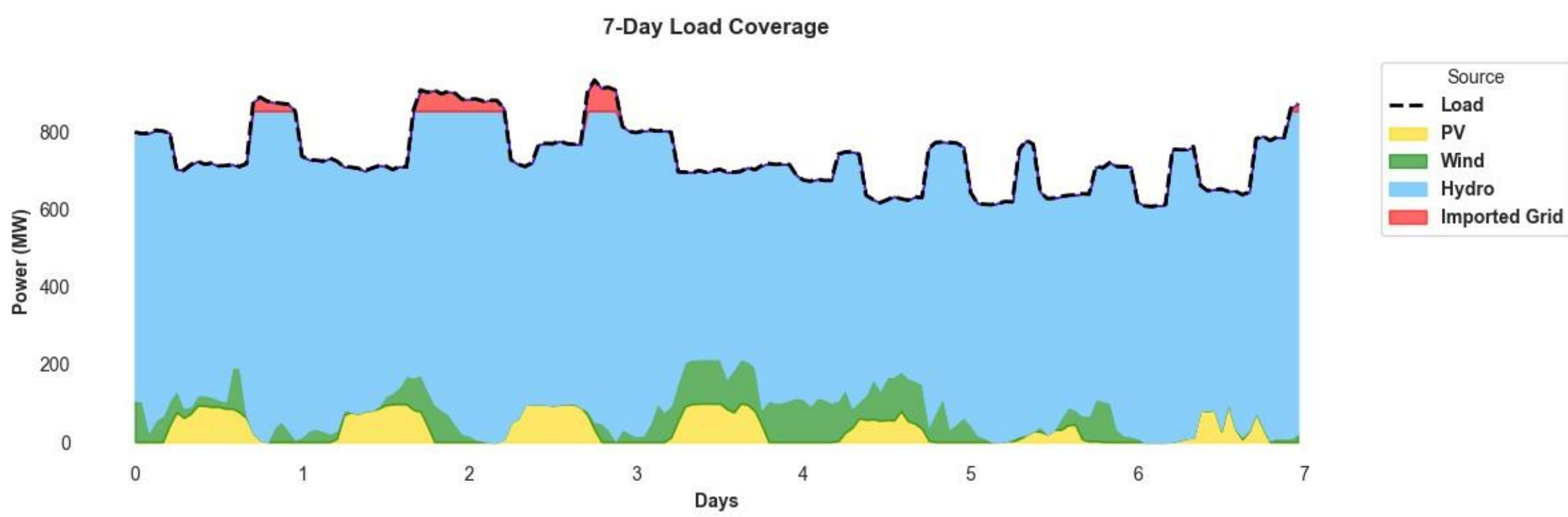

Figure 6: 7 days stacked profile at Dalles Dam

## Case-2 The Conowingo hydro power plant at Maryland

Due to geographic constraints near the Conowingo Hydropower Plant, there is limited land availability for deploying large-scale wind or photovoltaic (PV) systems. Therefore, a realistic PV capacity of approximately 130 MW was considered for integration. This PV plant is expected to provide partial grid support to meet the data center's energy demand. Figure 7 shows the 1 MW normalized solar profile at Maryland area for the month of May. Similarly, Figure 8 shows the 1 MW normalized wind profile at Maryland area. Figure 9 illustrates the 24-hour operational profiles, including the data center load, optimized hydro generation, PV output, and the additional grid power required to meet the residual demand. Because of the restricted capacity of renewable generation at this location, a portion of the energy must be imported from the external grid. Figure 10 compares the default (scheduled) hydro generation profile with the optimized hydro output. The plot highlights a significant shift in hydro dispatch strategy, aiming to better align generation with net load after accounting for PV contribution. Table 3 presents a detailed cost breakdown for the system, including capital expenditures (CAPEX) for PV and hydro, operational costs, and grid purchase costs. These values highlight the trade-offs between renewable integration, hydro optimization, and grid dependence. The Figure 11 shows the stacked plot of the different generation for 7 days. As shown in the plot due to the limitation of the renewable area land and an extra burden of 55 MW IT load from the data center there are additional reliance on the grid. Moreover the transformer/electrical infrastructure limit was kept at 400 MW.

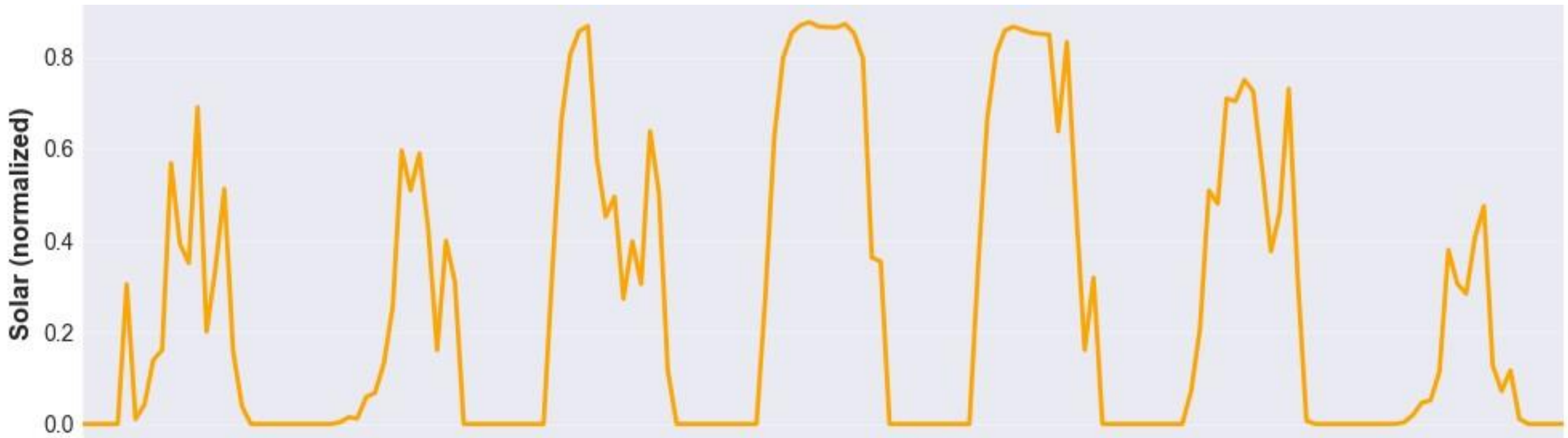

Figure 7: 1 MW solar PV profile at Conowingo Dam normalized

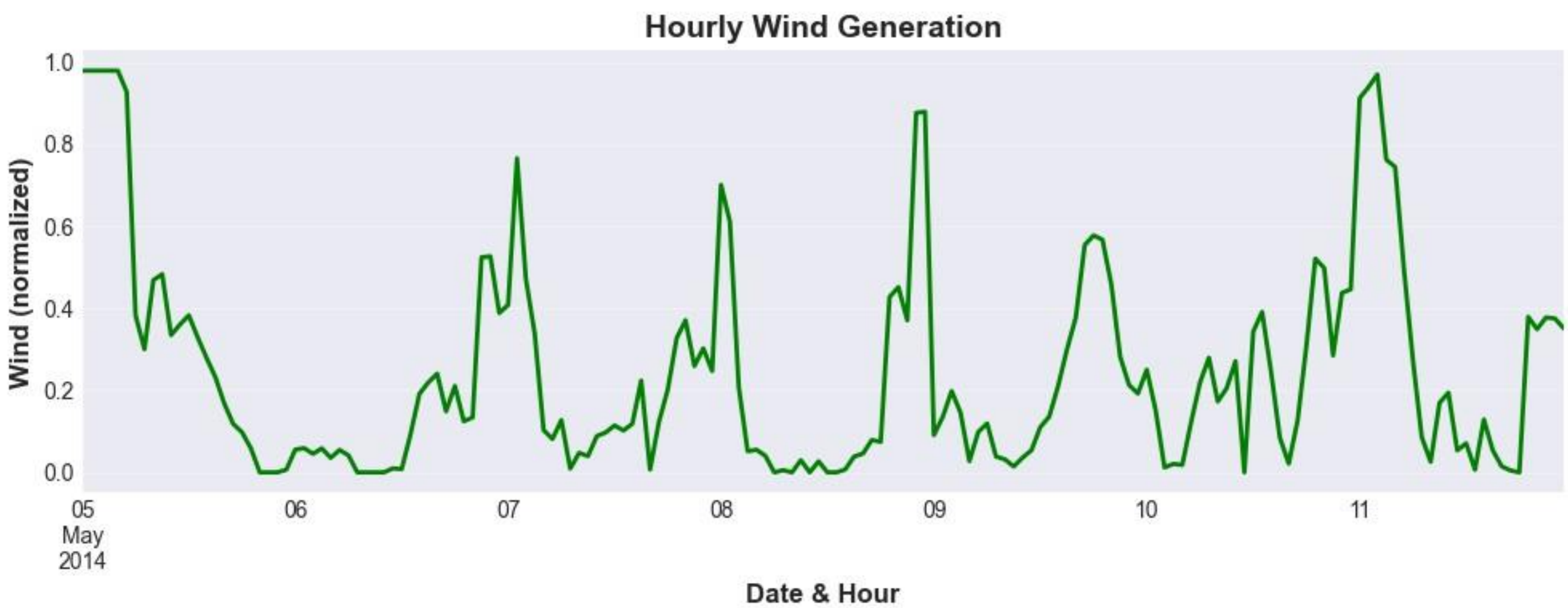

Figure 8: 1 MW wind profile at Conowingo Dam normalized

Table 3: System Cost Breakdown

| Component | Cost (USD) |
|---|---|
| PV CapEx + O&M | $219,219,000.00 |
| Wind CapEx + O&M | $0.00 |
| BESS CapEx + O&M | $0.00 |
| Grid Cost | $2,036,589.59 |
| Total Cost | $221,255,589.59 |

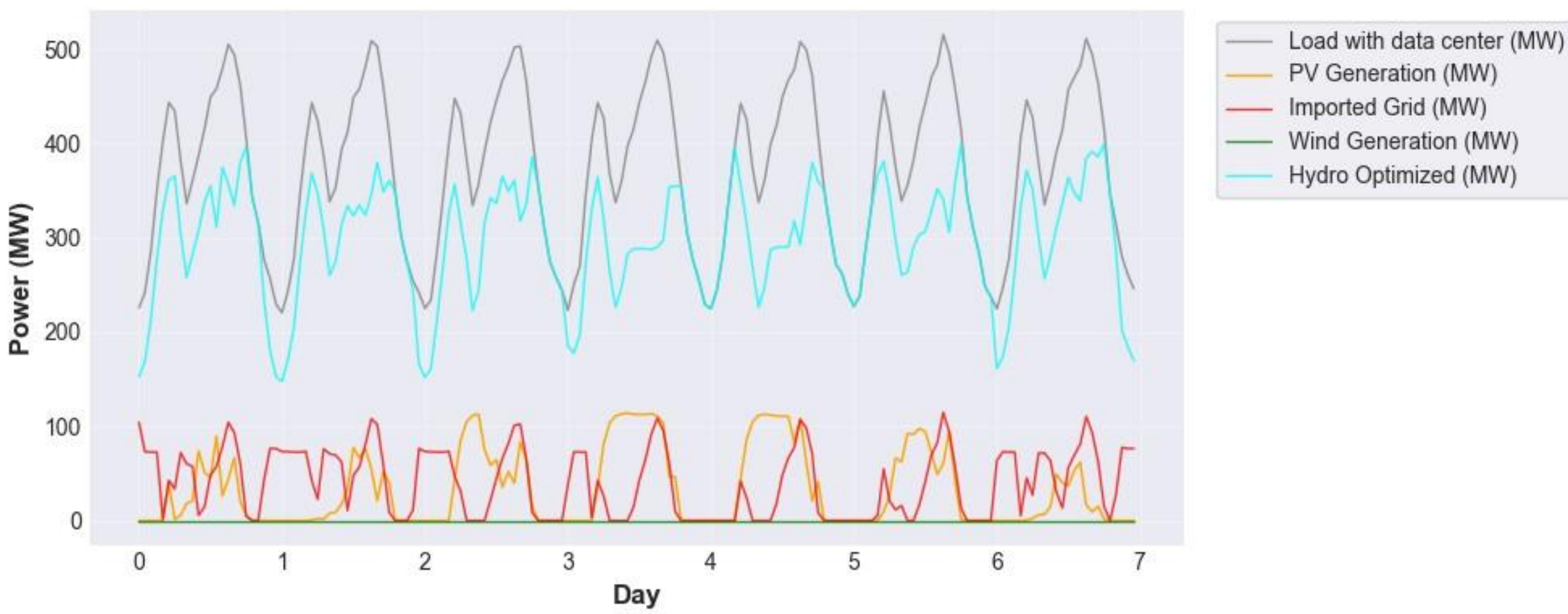

Figure 9: 7 days profile at Conowingo Dam

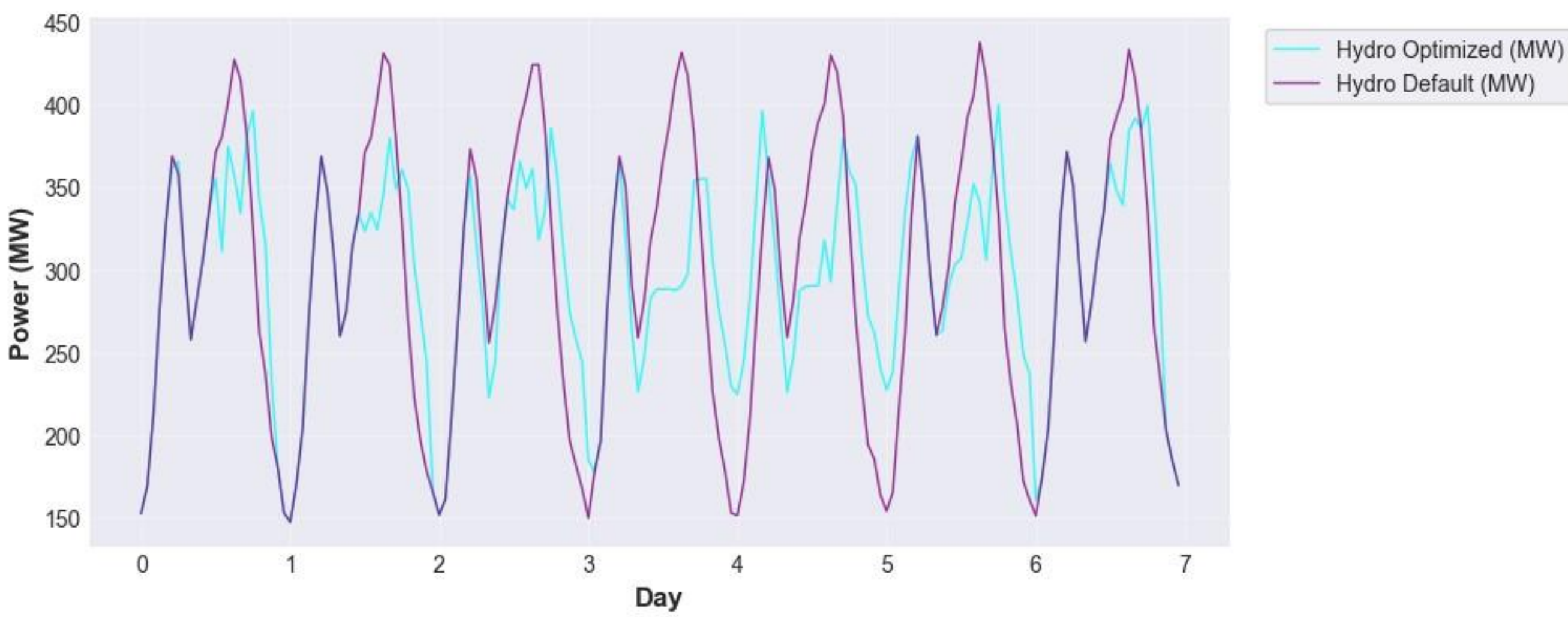

Figure 10: 7 days optimized hydro profile at Conowingo Dam

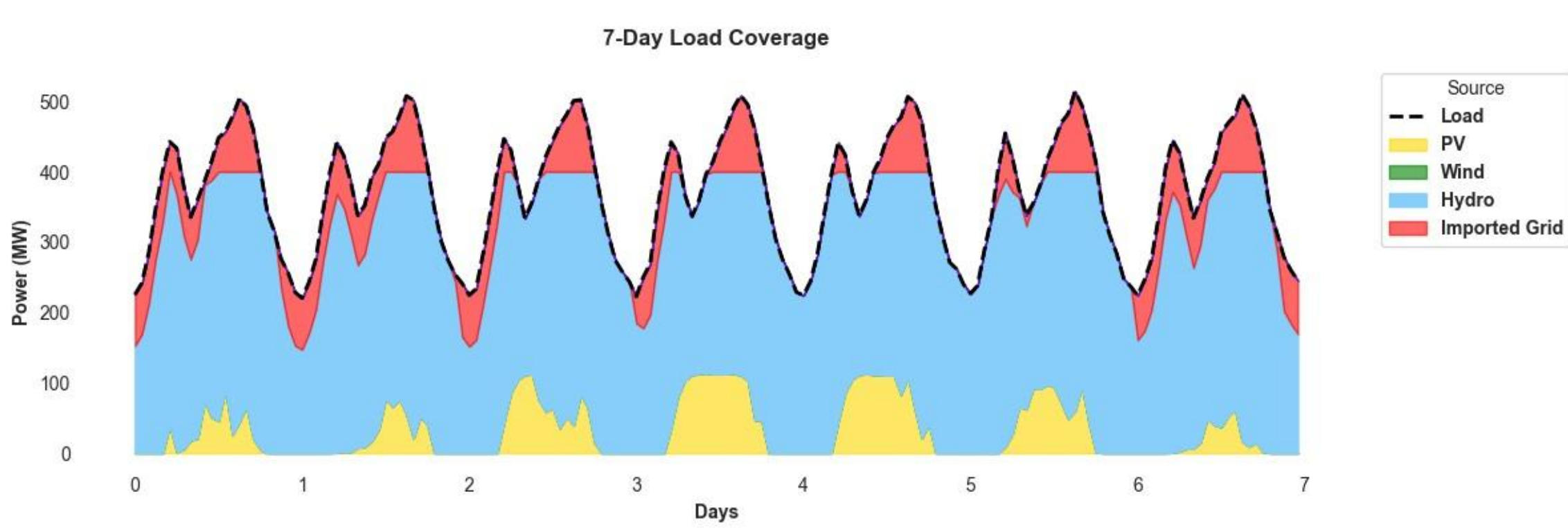

Figure 11: 7 days stacked profile at Conowingo Dam

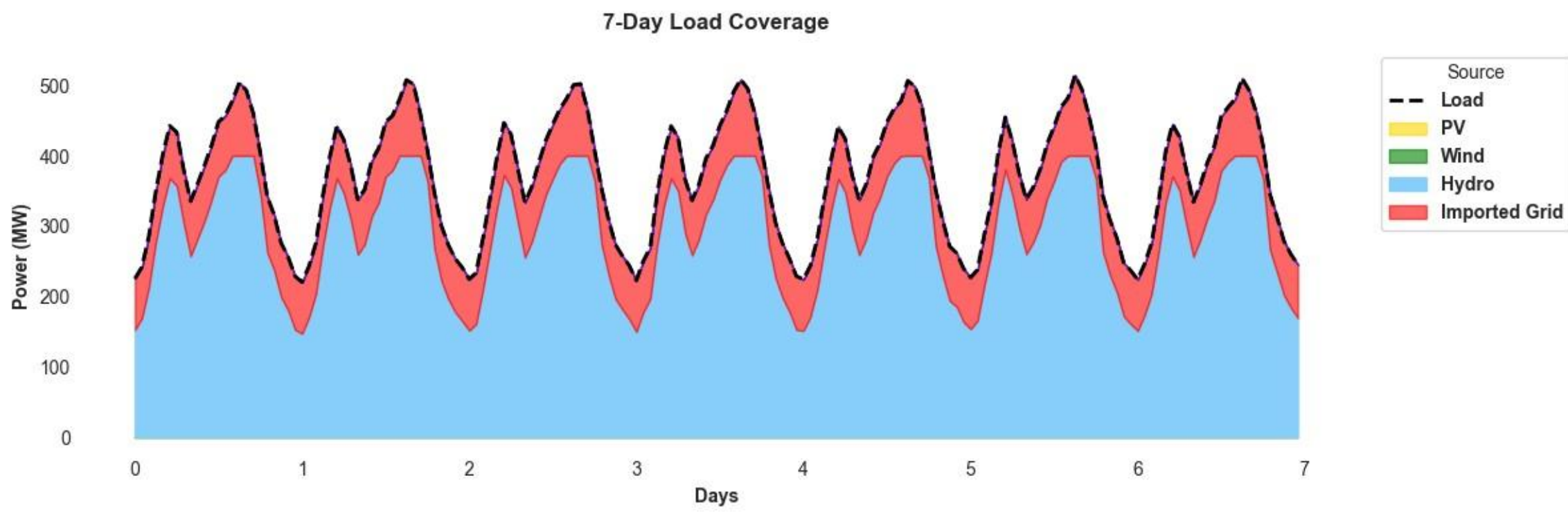

Figure 12: 7 days stacked profile at Conowingo Dam without renewable sources